\documentclass[pra,aps,12pt,groupedaddress,nofootinbib]{revtex4-2}  
\usepackage{graphicx}
\usepackage[nointegrals]{wasysym} %
\usepackage[export]{adjustbox}
\usepackage{amsmath,amsfonts,amssymb,latexsym}
\usepackage{hhline}
\usepackage{bm}
\usepackage{verbatim}
\usepackage{enumitem}
\usepackage{mathrsfs}
\usepackage{slashed}
\usepackage{empheq}
\usepackage{physics}

\usepackage{graphicx}
\usepackage{dcolumn}
\usepackage{bm}
\usepackage{multirow}

\usepackage[normalem]{ulem}

\usepackage{amsmath}
\usepackage{amsthm}
\usepackage{amstext}
\usepackage{amssymb}
\usepackage{mathrsfs}
\usepackage{amsfonts}
\usepackage{amsbsy} 

\usepackage{braket}

\usepackage[all,matrix,cmtip]{xy}
\usepackage{mathtools}

\usepackage{xcolor}
\definecolor{red}{rgb}{1,0,0}
\definecolor{blue}{rgb}{0,0,1}
\definecolor{dblue}{rgb}{0,0,0.4}
\definecolor{green}{rgb}{0,1,0}
\definecolor{black}{rgb}{0,0,0}
\definecolor{white}{rgb}{1,1,1}
\definecolor{pastelblue}{RGB}{20,93,160}

\definecolor{brn}{rgb}{.8,.4,.0}
\definecolor{redo}{rgb}{1,.5,.0}
\definecolor{ddgrn}{rgb}{0,0.4,0}
\definecolor{dgrn}{rgb}{0,0.55,0}
\definecolor{dbl}{rgb}{0,0,0.5}

\usepackage[colorlinks,citecolor=pastelblue,linkcolor=pastelblue,urlcolor=pastelblue]{hyperref}

\newcommand{\bpm}{\begin{pmatrix}}
	\newcommand{\epm}{\end{pmatrix}}
\newcommand{\bmm}{\begin{matrix}}
	\newcommand{\emm}{\end{matrix}}
\newcommand{\bvm}{\begin{vmatrix}}
	\newcommand{\evm}{\end{vmatrix}}

\usepackage{euscript}

\makeatletter
\newsavebox{\@brx}
\newcommand{\llangle}[1][]{\savebox{\@brx}{\(\m@th{#1\langle}\)}%
	\mathopen{\copy\@brx\kern-0.5\wd\@brx\usebox{\@brx}}}
\newcommand{\rrangle}[1][]{\savebox{\@brx}{\(\m@th{#1\rangle}\)}%
	\mathclose{\copy\@brx\kern-0.5\wd\@brx\usebox{\@brx}}}
\makeatother

\allowdisplaybreaks

\newcommand{\bs}{\boldsymbol}
\usepackage{booktabs}
\usepackage{tabularx}
\usepackage{array}

\begin{document}


\title{Fractionalized metals from doped anyons: Application to tMoTe$_2$}

\author{T. Senthil}
\email{senthil@mit.edu}
\affiliation{
Department of Physics, Massachusetts Institute of Technology,
Cambridge, Massachusetts 02139, USA
}%

\date{\today}
\begin{abstract}
 A fluid of mobile anyons may arise naturally when a Fractional Quantum Anomalous Hall (FQAH) state is doped. Motivated by recent experiments on twisted $MoTe_2$, we study metallic phases obtained by doping the $\sigma_{xy} = 2e^2/3h$ state. We propose that the high-resistivity metal observed adjacent to this FQAH state is a $Z_3$ Orthogonal Metal: a non-Fermi liquid with sharp charge $1/3$ fermionic quasiparticles coupled to a discrete $Z_3$ gauge field but with no sharp electronic quasiparticle. For a dilute gas of three species of charge-$1/3$ anyons with $\pi/3$ statistics, we construct two $U(3)$ symmetric $Z_3$ Orthogonal Metals, distinguished by whether the gapless fermions transform as $SU(3)$ triplets or singlets. We show that these states naturally yield large electrical resistivities even when the fractionalized quasiparticles are in a good metallic regime. Pairing of the charge-$1/3$ fermions produces an ordinary charge-$2e$ superconductor, smoothly connected to a BCS state but obtained through an intrinsically fractionalized normal state. We discuss experimental signatures of the idea that the normal metallic state in the lightly doped $2/3$ state may have fractionalized charge carriers. 
\end{abstract}
\maketitle

\section{Introduction} 
A fluid of mobile anyons may arise naturally when Fractional Quantum Anomalous Hall (FQAH) states (or more generally Fractional Chern Insulators (FCI)) are doped. The study of such fluids was initiated in the late 1980s and shown to provide a non-BCS mechanism for superconductivity~\cite{Laughlin1988_anyonSC,Lee1989_anyonSC,Fetter1989_anyonSC_RPA,Chen1989_anyonSC,Wen1990_anyonSC,Tang2013_anyonSC} - since dubbed `anyon superconductivity'\footnote{The term refers to a mechanism for stabilizing a superconductor through creation of  a fluid of anyons. The superconducting state itself is an electronic superconductor and may even be smoothly connected to a BCS state of electrons}. The discovery of the FQAH in moire materials~\cite{Cai2023_FQAHTMD,Park2023_FQAH_TMD,Xu2023_FQAHTMD,Zeng2023_FQAHTMD,Lu2023_FQAHPenta,Lu2025_EQAH} has led to  a revival~~\cite{Shi2024_doping,Divic2024_HofHubb, Kim2024_anyonSC,Zhang2025_SU(3)1dope,Pichler2025_anyonSC,Shi2025_dopeMR,Nosov2025_plateau,Shi2025_anyon_delocalization,Han2025_anyonexciton,Nakajima2025_thermo_anyon,Kuhlenkamp2025_HofHubb,Shi2025_nonAbelian_TSC,lotrivc2026phases,Fan2026_weakpairing_SC,Wang2026_U3anyonSC,shi2026superconductivity} in this topic with many papers expanding the basic ideas underlying the anyon superconductor but also demonstrating the occurrence of metallic phases~\cite{Shi2024_doping,shi2026superconductivity}. Amazingly experiments\cite{Xu2025_SCdopeTMD} on high quality twisted MoTe$_2$ show superconductivity developing at dopings near an FQAH state with Hall conductivity $\sigma_{xy} = \frac{2e^2}{3h}$. 

The bulk of the previous theoretical work has explored routes to superconductivity upon doping the $2/3$ state. Our focus in this paper instead is on a doped {\it metallic} ground state. Indeed a metallic state with a high residual resistivity of a few kOhms appears\cite{Xu2025_SCdopeTMD} as the first state adjacent to the FQAH at a lattice filling $\nu = 2/3 + \delta\nu$ in $tMoTe_2$. Superconductivity only appears after a non-zero doping $\delta\nu \approx 0.025$. Furthermore the superconductivity can be suppressed by a magnetic field and leads initially to a metallic ground state which seems to be the continuation of the low doping metal seen in zero field. Finally in the superconducting range of dopings, this metal has a longitudinal resistivity that increases as $2/3$ filling is approached as expected if its carrier density is determined by the doping $\delta\nu$ (rather than the deviation from the band edge $1-\nu$). Thus we propose to think of this metallic state as the `normal' state out of which the superconductivity develops, and further that it is a metal formed by charges doped into the $2/3$ state which naturally are anyons. 

In a recent study on the physics of a fluid of mobile charge $1/3$ anyons, the author, together with Z.D. Shi,  described\cite{shi2026superconductivity} the possibility of a non-fermi liquid metal that emerges at low density. This metal has a Fermi surface of sharply defined charge 1/3 fermions coupled to a discrete $Z_3$ gauge field. Despite the presence of these charge $1/3$ quasiparticles, the electron quasiparticle does not exist at low energies, and hence the metal is a non-Fermi liquid. We denoted this state a $Z_3$ Orthogonal Metal ($Z3OM$). Closely similar states were  first introduced and studied in Ref. \cite{nandkishore2012orthogonal}, though not in the context of anyon fluids. The terminology Orthogonal Metal indicates that though there are well-defined quasiparticles at low energies in this metal, they are orthogonal to the physical electron, {\it i.e}, they have zero overlap. 

The purpose of this paper is to elaborate on the $Z3OM$ phase in the context of the doped $2/3$ FQAH state. We will construct two distinct such phases, describe their phenomenology, and show how they provide explanations of many features of the non-superconducting ground state of the lightly doped $2/3$ state. In particular we propose that the observed metal is not a dirty electron metal, but a relatively clean metal of fractionalized charge-1/3 quasiparticles. Its large resistivity arises because electrical conductivity is suppressed by the square of the fractional charge. 

 These $Z3OM$ states are natural symmetry-preserving metallic candidates for the ideal or nearly ideal anyon gas, though determining the true ground state requires an energetic calculation beyond the present work. 

We describe the properties of the superconducting state that develops by pairing out of the Z3OM states, and show that it is smoothly connected to a BCS charge $2$ superconductor (SC). We discuss a number of experimental consequences of the charge fractionalization in the `normal' metallic state that can provide a clear test of our proposal.  These include fractional Josephson oscillations in $SC -Z3OM-SC$ junctions, $(2e/3)$ shot noise at $SC-Z3OM$ interfaces, distinctive quantum oscillation slopes, suppressed electron tunneling, and an enhanced Wiedemann-Franz ratio. We also show that at low doping, in the presence of disorder, the superconductor is likely to be an Anomalous Vortex Glass\cite{Shi2025_anyon_delocalization} in a disordered sample, which may be consistent with the observed gradual development of the superconductor in experiments\cite{Xu2025_SCdopeTMD}. 

\section{Models for doped charge $1/3$ anyons}
We consider the possibility that near the $2/3$ state, the dopants are charge $1/3$ anyons.  As emphasized in earlier papers\cite{Shi2024_doping,Shi2025_anyon_delocalization,Nosov2025_plateau,Fan2026_weakpairing_SC,shi2026superconductivity}, microscopic lattice translation symmetry acts projectively on the anyons of the $2/3$ state. This leads to 3 dispersion minima for the $1/3$ charged anyon. At small doping, there will therefore be 3 species of these anyons which are related by the action of microscopic lattice translation symmetry. We are thus lead naturally to consider a model of 3 species of the $1/3$ anyons at a non-zero density. These anyons have statistics $- \pi/3$. In what follows it is notationally convenient to instead consider doping the $1/3$ state with charge $1/3$ anyons with statistics $+ \pi/3$. Results for the doped $2/3$ state can be readily obtained by particle-hole transformation, which will require adding the response of a filled Chern band.

We can describe the statistics $\pi/3$ anyons as bosons with attached flux that implements the fractional statistics. Thus we consider a Hamiltonian defined in a Hilbert space of symmetric ({\it i.e} bosonic) many-body wavefunctions 
\begin{eqnarray}
    H & = & H_0 + H_{int} \\
    H_0 & = & \sum_{i = 1}^N \sum_{I = 1}^3 \frac{\left( \bs{p}_{iI} - \bs{b}_{i}\right)^2}{2m} \\
    H_{int} & = & \frac{1}{2} \sum_{i\neq j} \sum_{I} V_0 (\bs x_{iI} - \bs{x}_{jI} ) + \frac{1}{2} \sum_{ij} \sum_{I \neq J} V_1(\bs{x}_{iJ} - \bs{x}_{jJ}) 
\end{eqnarray} 
together with the constraint 
\begin{equation} 
3\bs{\nabla} \times \bs{b}(\bs{x}) = 2\pi \sum_i \sum_I \delta^{2}(\bs{x} - \bs{x}_{iI}) 
\end{equation} 
Here $\bs{b}$ is the spatial part of a $U(1)$ gauge field that implements the flux attachment. $V_0$ and $V_1$ are same-species and inter-species two body interactions which we will take to be repulsive.

To begin with let us consider a microscopic system with short-ranged repulsive interactions between the electrons. Then the anyons will also only have short ranged interactions. Of course the effects of particle statistics is infinitely long-ranged, and will be treated as such.

A single anyon is not a point object but rather has a size $l_a$ that is determined by the short distance physics of the parent FQAH. In the lowest Landau Level or Aharanov-Casher band realization, numerical estimates ( see, for eg, Refs. \cite{liu2015characterization,Yan2025_anyondisp}) suggest (for the Laughlin $1/3$ state) $l_a \approx 2-3 l_B$ where $l_B$ is the magnetic length. Then the range of the interactions $V_{0,1}$ is at least $l_a$.  Consider the limit of low anyon density $n$ in the sense that  $n l_a^2 \ll 1$. Then the inter-anyon separation is much larger than the minimum range of their interaction. Further, as anyon wavefunctions will decay as two anyons approach each other (unlike bosons), we might expect that the leading order physics in a low density expansion is captured by the ``ideal anyon" model defined by ignoring the $H_{int}$ terms. As explained below, the ideal anyon model has an exact $U(3)$ global symmetry. This includes a $U(1)$ factor corresponding to conservation of total anyon number and an $SU(3)$ factor that rotates between the three different species. 

At higher densities, as the anyons start overlapping,  their interactions become important\footnote{Given the expected large anyon size, this will happen at very low densities.} We will restrict to considering two body repulsive  interactions. They are naturally organized in terms of their transformations under the $U(3)$ symmetry. A $U(3)$-singlet interaction corresponds to one between the total densities of the three species. We also expect $U(3)$ breaking interactions which can be written as an interaction between $SU(3)$ pseudospins of the 3 fermion species.  

The long range part of a Coulomb interaction is readily incorporated into the model as it involves the total charge density and hence is $U(3)$ symmetric.

As usual, we can use a Chern-Simons description of this system with the (imaginary time) action 
\begin{equation} 
\label{eq: bosonrep}
{\cal S} = \int d^2x d\tau \sum_I \bar{\Phi}_I \left(i\partial_t + b_0 + \frac{1}{2m}(-i \vec \nabla - \vec b)^2 \right)\Phi_I - \frac{3}{4\pi} bdb + \frac{1}{2\pi} Adb + {\cal S}_{int}
\end{equation} 
where the $S_{int}$ term descends from the interaction $H_{int}$. 
Here $b$ is a dynamical $U(1)$ gauge field, and $A$ is a background gauge field that probes the global $U(1)$ symmetry. If we set $S_{int} = 0$, {\it i.e}, in the ideal gas limit, then the action has manifest global $U(3)$ (apart from the $U(1)$) symmetry). The bosons $\Phi_I$ transform in the fundamental representation of $SU(3)$. The charge conservation is implemented through the conservation of $\bs{\nabla} \times \bs{b}$. 

$S_{int}$ will include both $SU(3)$ preserving as well as $SU(3)$ breaking repulsive interactions, though the latter may be of smaller strength\cite{shi2026superconductivity}. A $Z_3$ `clock' symmetry is guaranteed by the microscopic lattice translation symmetry. 

\section{Two $Z_3$ Orthogonal Metals}
At non-zero charge density, the bosons see a mean magnetic flux such that the bosons are at a total filling $\nu_t = 3$. Let us consider $SU(3)$ symmetric mean field states. To that end we consider a parton description $\Phi_I = \Psi_I d$ where $\Psi_I$ is an $SU(3)$ fundamental fermion and $d$ is an $SU(3)$ singlet fermion. As usual this representation introduces a $U(1)$ gauge redundancy. We can contemplate two distinct ground states as explained below. 

(i) {\it $SU(3)$ triplet metal}: We let $d$ see the magnetic field associated with $b$ while $\psi_I$ form 3 Fermi surfaces. Then $\nu_d = 3$ and $d$ can fill 3 Landau levels. At long length scales we then get an effective theory 
\begin{equation}
{\cal L} = {\cal L}[\psi_I, a] + \frac{3}{4\pi} (a-b) d(a-b) + 6CS_g  - \frac{3}{4\pi} bdb + \frac{1}{2\pi} Adb 
\end{equation} 
where $a$ is a dynamical $U(1)$ gauge field\footnote{As explained in this context in previous papers~\cite{Shi2024_doping} it is helpful as a bookkeeping device to place the theory on a general oriented space-time with a metric $g$ and keep track of distinctions between spin$_c$ connections and $U(1)$ gauge fields. Then $a$ is actually a spin$_c$ connection (fields with odd charge under $a$ are fermions). The $CS_g$ is a gravitational Chern-Simons term that enables keeping track of edge modes.}.  
Clearly the self Chern-Simons terms for $b$ cancel and we get 
\begin{equation}
\label{eq: z3omt}
{\cal L}^+_t = {\cal L}[\psi_I, a] + \frac{3}{4\pi} ada  + \frac{1}{2\pi} bd (A - 3a) + 6CS_g
\end{equation} 
This is a $Z_3$ Orthogonal Metal with 3 Fermi surfaces formed by an $SU(3)$ fundamental fermion. The fermions carry charge-$1/3$ and are coupled to a (twisted) $Z_3$ gauge theory. The discrete gauge field implies that the low energy gapless fermions near the Fermi surface can be described through Fermi liquid theory. The electron quasiparticle itself will however not be well-defined as it is a composite of three low energy fermion quasiparticles.  Precisely this state was obtained in Ref. \cite{shi2026superconductivity} through a different (Jain) composite fermion representation of the 1/3 anyons. Here we will denote it as $Z3OMt$ where the $t$ is a reminder that the gapless fermions are $SU(3)$ triplets. At  anyon density $n$, the physical charge density $\delta \rho = \frac{n}{3}$ ($\delta \rho$ is defined by the deviation from the FQAH filling). Then each of the three Fermi surfaces have a Fermi momentum  $K_{FI} = \sqrt{4\pi \delta \rho}$. 

(ii) {\it $SU(3)$ singlet metal}: Alternately we let the $\psi_I$ see the magnetic field associated with $b$ while $d$ forms a Fermi surface. Each $\psi_I$ is then at $\nu_I = 1$ and can form an integer quantum Hall state. The resulting ground state of 3 filled Landau levels of the $\psi_I$ is $SU(3)$ invariant. This leads to the Lagrangian 
\begin{equation} 
\label{eq: z3oms}
{\cal L}^+_s = {\cal L}[d,a] + \frac{3}{4\pi} ada  + \frac{1}{2\pi} bd (A - 3a) + 6CS_g
\end{equation}
This is also a $Z_3$ orthogonal metal of charge $1/3$ fermions which are $SU(3)$ singlets. We will denote it $Z3OMs$. Note that the global $SU(3)$ has trivial action in the bulk in this state. However there will be gapless $SU(3)$ edge currents associated with the integer quantum Hall state of the $\psi_I$ fermions. In Appendix \ref{app: z3omsJain} , we show how to construct this $Z3OMs$ state within the composite fermion formalism of Ref. \cite{shi2026superconductivity}. As there is just a single Fermi surface in $Z3OMs$, it will have a Fermi momentum $K_F = \sqrt{12\pi \delta \rho}$. 

Both these are metallic non-fermi liquid states (in the sense that the electron quasiparticle is not well-defined though the charge $1/3$ fermions are sharp quasiparticles). Either of these are candidate $U(3)$ invariant ground states for the anyon model with a $U(3)$ symmetric Hamiltonian. This includes the ideal anyon limit. Breaking $SU(3)$ weakly will however not destabilize these states. 

In applying to the  $-\pi/3$ statistics anyon fluids appropriate to the doped $2/3$ FQAH state, the required particle-hole transformation changes Eq. \ref{eq: z3oms} to 
\begin{equation} 
\label{eq: z3omsph}
{\cal L}^-_s = {\cal L}[d,a] - \frac{3}{4\pi} ada  - \frac{1}{2\pi} bd (A - 3a) + \frac{1}{4\pi} AdA -4CS_g
\end{equation}
(The filled Chern band contributed $\frac{1}{4\pi} AdA + 2CS_g$ which explains the last two terms above). A similar change also applies to Eqn. \ref{eq: z3omt}. We use the superscript $\pm$ to denote the sign of the $\pm \pi/3$ statistics and the subscript $t/s$ to denote whether it is the triplet or singlet OM\footnote{The $Z3OMs$ state has also been constructed by Zhaoyu Han and Eslam Khalaf (private communication).}. 

Recently Ref.~\cite{Fan2026_weakpairing_SC} proposed a simple charge $2$ superconducting ground state for the same ideal anyon model. The $SU(3)$ symmetry of the ideal anyon mode was not noticed in Ref. \cite{Fan2026_weakpairing_SC}. As we discuss later, this - and other - superconducting states can be $SU(3)$ invariant. However the path to such a superconductor in that paper relies on a mechanism that is not manifestly $SU(3)$ invariant, and hence it is not clear to us if the mechanism identified there applies to the ideal anyon model.  In particular the `parent' metallic state out of which the SC develops through a pairing instability is, if treated in weak coupling as in Ref.~\cite{Fan2026_weakpairing_SC}, not $SU(3)$ invariant.  Nevertheless the discussion of Ref. \cite{Fan2026_weakpairing_SC} can apply to interacting anyon models where the $V_0$ interaction dominates over the $V_1$ interaction, and the $SU(3)$ symmetry is explicitly broken.

\subsection{Regime of validity} 
In applying the low energy effective theory (Eqn. \ref{eq: z3omt} or Eqn. \ref{eq: z3oms}) of either $Z3OM$ state to the doped $1/3$ FQAH state, we face a puzzle. If the density of anyons goes to zero, then these theories reduce to a twisted (Dijkgraaf-Witten) $Z_3$ gauge theory rather than to the familiar $U(1)_3$ theory of the undoped $1/3$ state. The resolution of this puzzle is that Eqn. \ref{eq: z3omt} or Eqn. \ref{eq: z3oms} only applies at length scales large compared to the inter-anyon separation $1/\sqrt{n}$, and hence cannot be continued to the zero density limit. This is because in obtaining these effective theories, we integrated out the gapped fermions that form integer quantum Hall states. This is valid at length scales larger than their magnetic length $1/\sqrt{B_{eff}}$ where $B_{eff}$ is the self-consistent magnetic field they move in. However as $B_{eff} \propto n$, their regime of validity is at length scales larger than $1/\sqrt{n}$. 

\subsection{ Stability to interactions} 
So long as the interactions between the emergent fermions is repulsive in either of the two $Z3OM$ states, they will be locally stable - upto Kohn-Luttinger effects leading to weak superconductivity which will be weak at small density. They could however become important with increasing density. Note that any interaction that involves the total charge density $\sum_I \bar{\Phi}_I \Phi_I$ is $U(3)$ invariant. Thus the long wavelength part of a microscopic density-density interaction will lead to a $U(3)$ symmetric repulsive interaction between the low energy fermions, and will lead to the usual Fermi liquid renormalizations and interactions of the low energy fractionalized quasiparticles. These states will also be stable to weak $SU(3)$ non-symmetric repulsive interactions. For $Z3OMs$, this is obvious as the $SU(3)$ simply does not act on the low energy theory. For $Z3OMt$, much like in usual Landau Fermi liquid theory, such $SU(3)$ non-symmetric repulsive interactions will modify the detailed Fermi liquid interactions but will not destabilize the state. 

In what follows we will explore the possibility that either of these states is a `normal' metallic state for the doped $1/3$ FQAH state. In applying to experiments near the prominently observed $2/3$ state, we will need to consider a gas of statistics $-\pi/3$ anyons, and will use the theory in Eqn. \ref{eq: z3omsph} or its equivalent for $Z3OMt$.   

\section{Transport and application to $tMoTe_2$}
We now explore the possibility that either of these metallic states is the resistive metallic state seen at lattice filling $\nu = 2/3+ \delta\nu$ ($\delta \nu > 0$) in tMoTe2. We will see that these states have many attractive features that may explain the phenomenology of this resistive state. Further we present evidence that this resistive state is the `normal' state out of which superconductivity eventually emerges as $\delta\nu$ is increased. Thus one of the $Z3OM$ states presents an attractive candidate normal state for lightly doped tMoTe2 at lattice filling $\nu = 2/3 + \delta \nu$. 

Let us describe key features of the experimental data of Ref. \cite{Xu2025_SCdopeTMD}. Upon doping the $2/3$ state toward lattice filling $1$, a metallic region with a large residual resistivity is first seen in a narrow range of doping. Increasing the doping further leads to a dome-shaped superconducting region till $\nu \approx 0.75$ (at low displacement field). Beyond that a metallic ground state results which persists all the way upto full band filling. 

Our main focus will be on the metallic resistive state seen close to the $2/3$ state. This metallic region appears as a peak in the resistivity in between the FQAH and the SC which both have zero $\rho_{xx}$. It is seen in multiple devices, persists to high temperature (much higher than the superconducting $T_c$) and its width in doping increases with increasing $T$. At low-$T$, the peak width saturates to a constant $\Delta \nu \approx 0.022$ while the peak height is $\approx 10$ kOhms\footnote{In our earlier work\cite{Shi2025_anyon_delocalization}, as well as Ref. \cite{Nosov2025_plateau}, this peak was interpreted as the universal resistivity peak associated with the FQAH-SC transition. However the saturation of the peak width to a non-zero constant at low $T$ is problematic in that interpretation.}. 
Interestingly when the superconductivity is suppressed by a magnetic field, the width of this metallic region expands to occupy the entire doping range where SC is seen at $B = 0$. Indeed suppressing the SC with a field of $0.9 T$ leads to a metallic state with a large $\rho_{xx} \approx 6$ kOhms (Fig 4 in Ref. \cite{Xu2025_SCdopeTMD}).

In light of these facts, we view this metallic state as the parent `normal' state out of which superconductivity develops. 
Below we will explore an interpretation of this state as either of the two $Z3OM$ states described above. For concreteness we focus on the $Z3OMs$ state though the results for $Z3OMt$ will essentially be similar. 
Let us consider transport in this state. We will assume that the $d$ fermions have an elastic mean free path $l$ coming from impurities or other sources of disorder. In Eq. \ref{eq: z3omsph} we can discuss transport by setting $a = A/3$ ({\it i.e} by assigning charge $1/3$ to these fermions).

We begin with the Hall conductivity. The term $-\frac{1}{4\pi}  ada$ then leads to a contribution to the Hall conductivity which will combine with the filled Chern band contribution $\frac{1}{4\pi}AdA$ to give $\sigma_{xy} = \frac{2e^2}{3h}$. In addition the $d$ band may have a Berry curvature near the band bottom (which is allowed by symmetry), which will lead to a Fermi surface anomalous Hall effect. Let ${\cal B}_0$ be the Berry magnetic field (determined by microscopic details) at the $d$-band bottom. The $d$-Fermi surface contribution to the Hall effect is then ${\cal B}_0 \pi k_F^2/(2\pi) = 6\pi {\cal B}_0 \delta \rho$. Thus we can write the full Hall conductivity as 
\begin{equation} 
\sigma_{xy} = \frac{e^2}{h}\left (\frac{2}{3} + 6\pi{\cal B}_0 \delta \rho \right) 
\end{equation}

The longitudinal conductivity of the charge $1/3$ fermions is simply (assuming a density independent mean free path $l$)
\begin{equation} 
\sigma_{xx} = \frac{e^2}{18h} k_F l 
\end{equation} 
The factor of $1/18$ has two origins: the factor $1/9$ comes from the squared fractional charge $q^2/e^2=1/9$, while the remaining factor $1/2$ is the standard spinless two-dimensional Drude factor for a single circular Fermi surface, $\sigma_{xx}=(q^2/h)(k_Fl/2)$. Thus the small $\sigma_{xx}$ does not by itself imply a bad metal; the relevant semiclassical parameter remains $k_Fl$\footnote{Note also that the small $\sigma_{xx}$ does not lead to Anderson localization so long as  $K_Fl$ is not too small.}.  The resistivities follow immediately: 
\begin{eqnarray}
    \rho_{xx} & = & \frac{h}{e^2} \frac{\frac{k_Fl}{18}}{\left(\frac{k_Fl}{18}\right)^2 + \left(\frac{2}{3} + 6\pi {\cal B}_0 \delta \rho\right)^2 } \\
    |\rho_{xy}| & = & \frac{h}{e^2} \frac{\frac{2}{3} + 6\pi {\cal B}_0 \delta \rho}{\left(\frac{k_Fl}{18}\right)^2 + \left(\frac{2}{3} + 6\pi {\cal B}_0 \delta \rho\right)^2 }
\end{eqnarray}
with $k_F = \sqrt{12\pi \delta \rho} = \frac{1}{a_M}\sqrt{\frac{24 \pi }{\sqrt{3}} \delta\nu}$ where the last expression is in terms of the moire lattice spacing $a_M$ and the deviation $\delta \nu > 0 $ of the filling from $2/3$. To emphasize the basic features,  let us set ${\cal B}_0 = 0$ and  note that at $\delta \nu = 0.06$  the resistivities are $\rho_{xx} \approx 0.7 h/e^2$, $\rho_{xy} \approx 0.5 h/e^2$.  We see that it is straightforward to obtain fairly large $\rho_{xx}$ and $\rho_{xy}$ values even in a good metallic regime (measured by $k_F l \approx 16$).

Note that as $\delta \nu$ increases, $k_F$ quickly becomes a sizeable fraction of the linear Brillouin zone size. Thus the approximation of the quadratic dispersion of the charge 1/3 fermions will only be valid at small $\delta \nu$. The mean free path that enters here is that of the charge $1/3$ fermions. For smooth disorder that couples to the charge density, the scattering rate of these quasiparticles will be reduced compared to electrons due to their smaller electric charge. Thus we expect the mean free path of the quasiparticles to be enhanced compared to electrons. A detailed modeling of  disorder effects and estimation of the mean free path is beyond the scope of this paper (and will depend on modeling the dominant source of disorder in the devices used in the experiments). The Berry magnetic field ${\cal B}_0$ is also not currently known. Thus we will not attempt quantitative fits to the existing data based on these formulas.

An interesting feature of the observed superconductor is its behavior in a magnetic field. The data in Ref. \cite{Xu2025_SCdopeTMD} shows that $T_c$ is suppressed in a manner very reminiscent of standard BCS superconductors obtained through pairing a Fermi surface. However the field induced normal state has a large $\rho_{xx} \approx 6$ kOhms at $\nu = 0.74$, which further increases as the doping is reduced consistent with the idea that the mobile charge density is determined by deviation from $2/3$. The $Z3OM$ `normal' state allows us to have the SC emerge out of pairing quasiparticles at a `small' Fermi surface of doped carriers rather than of holes doped into the integer Chern insulator.  

Finally given the expectation that the superconductor is unconventional (not $s$-wave), there is no Anderson theorem protecting it from pair-breaking disorder. Thus we expect that the superconductivity is highly sensitive to disorder, as observed in experiments\cite{Xu2025_SCdopeTMD}.

\section{Pairing and superconductivity} 
It is interesting to ask about the fate of either $Z3OM$ state when the low energy charge $1/3$ fermions pair and condense. 
This will happen if there is an attractive interaction between these fermions. We will not specify the origin of such an interaction which can arise through many mechanisms. One possibility is the usual Kohn-Luttinger (KL) effect, particularly for the $Z3OMt$ state which has 3 symmetry related Fermi surfaces in the 3 valleys. As discussed in the literature on multi-valley systems, the KL effect can be enhanced by inter-valley particle-hole fluctuations. Thus with increasing doping density an effective attraction between the charge 1/3 fermions may develop and lead to pairing. Our focus in this section will be on describing the resulting state. 

First, the pairing will break the global $U(1)$ symmetry and lead to superconductivity. Second, the condensing Cooper pair is a charge $2/3$ boson. This pair condensation removes all the anyons in the theory so that the superconductor is SC and not SC$^*$ ({\it i.e} does not have coexisting intrinsic topological order).  We prove this formally in Appendix \ref{app: SC}. To see this physically we note that the anyons of the twisted $Z_3$ gauge theory that the fermions couple to can be labeled $(m_1, m_2)$ with $m_{1,2} = 0, 1, 2$. These anyons carry electric charge $m_1/3$. The fermion is the $(1,1)$ excitation, and the condensing boson is $(2,2)$. The only anyons that braid trivially with $(2,2)$ (and hence survive their condensation) are anyons with labels $(m,m)$. The anyons with $m_1 \neq m_2$ however determine the flux quantization in the resulting superconductor. Thus (for the theories near the $2/3$ state originating from Eqn. \ref{eq: z3omsph} or its triple analog) as $(m_1,m_2)$ has braiding phase $4\pi (m_2 - m_1) /3 \mod 2\pi$ around $(2,2)$, it is bound to a $4\pi (m_1 - m_2)/3 \mod 2\pi$ phase winding of the charge $2/3$ pair condensate. This implies that the phase of the physical Cooper pair winds by $4\pi (m_1 - m_2) \mod 6\pi$.  Hence the vortices all have physical Cooper phase wind by multiples of $2\pi$ despite the condensation of charge $2/3$ pairs, and the flux quantization is in units of $h/2e$. 
Note in particular that the $(1,0)$ anyon transmutes into a $- h/2e \mod 3h/2e$ vortex.  Thus we have a charge $2e$ SC obtained through an unconventional route. The physical electric charge is no longer a good quantum number in the superconductor, and hence the unpaired fermions (the $(1,1)$ particles) simply become the Bogoliubov quasiparticles of this superconductor. 

Thus pair condensation of the charge 1/3 fermions of these $Z3OM$ states leads to a charge-$2$ BCS superconductor. The chiral central charge can be any odd half-integer at weak pairing. For $Z3OMs$, suppose we pair into a an angular momentum $l$ channel (with $l$ odd). To determine the net chiral central charge we should add the contribution from the background $CS[A,g]$ terms in Eqn. \ref{eq: z3omsph} to get $c_- = -2 + l/2$. Note that as the $SU(3)$ does not act on the gapless bulk fermions of the $Z3OMs$, all these superconductors can be $SU(3)$ symmetric if that symmetry is present microscopically. 

For the $Z3OMt$ state, it is interesting to consider symmetric intervalley p + ip pairing. This yields a net $c_- = -2 + 3/2 = - 1/2$.  

Finally we note that - though the $SU(3)$ does not act on the $d$ fermions - for the SC obtained by pairing out of the $Z3OMs$ state, the basic $2\pi$ vortex will carry $SU(3)$ quantum numbers. This is because in the effective field theory describing this state, the $(1,0)$ anyon transforms in the fundamental of $SU(3)$. As we have seen, it transmutes into the $2\pi$ vortex in the superconductor, and hence the vortex will also transform in the fundamental. 

\section{Phase diagram} 
It is instructive to consider the possible phase diagram when the $2/3$ FQAH is doped toward $\nu = 1$. For dopings $\nu = 1- \delta'$ with $\delta' > 0$ and small, the most natural ground state is an ordinary Fermi liquid of holes with a Fermi surface area proportional to $\delta'$. For dopings $\nu = 2/3 + \delta$ with $\delta > 0$ and small, we have proposed that one of the $Z3OM$ states may be the ground state. Let us consider the ground states in the absence of superconductivity, for example, in a small magnetic field just enough to suppress the superconductivity (but not stabilize other competing phases). Then a simple possibility is that, as a function of chemical potential, there is a first order transition between a $Z3OM$ metal and the hole Fermi liquid. The charge density may jump across this transition. Thus if we consider the phase diagram as a function of filling, we will have an intermediate phase separated region with puddles of the hole Fermi liquid and the $Z3OM$ metal. 
Note however that this phase-separation scenario is not required for the existence of either of the $Z3OM$ states 
but provides a simple way to connect the low-doping fractionalized metal to the high-doping hole Fermi liquid.

The puddle formation due to the putative first order transition between the $Z3OM$ state and the hole Fermi liquid  might be the explanation of the large  resistivity seen beyond the superconducting dome. In that region quantum oscillations are also seen above $\approx 3 T$ (despite the few kOhm resistivity at $B = 0$) consistent with the hole Fermi surface state.  It is possible that the oscillations originate from Landau quantization within the hole Fermi surface puddles.

If the magnetic field is reduced, superconductivity may develop out of the $Z3OM$ state for $\delta$ large enough that the anyon interactions start mattering. 

There are other competing metallic states that may be stabilized in neighboring regions of the phase diagram. Based on the Jain composite fermion description, Ref. ~\cite{Shi2024_doping} described a period-3 Charge Density Wave (CDW) Fermi liquid metal with a background Integer Quantum Hall response at low doping. This was invoked in Refs. ~\cite{Shi2025_anyon_delocalization,Nosov2025_plateau} to interpret the observed Reentrant Integer Quantum Hall effect in $tMoTe_2$ as due to localization of the electronic quasiparticles of this metal.

\section{Experimental tests} 
We have argued that the $Z3OM$ state is theoretically plausible, and that it has transport properties broadly consistent with experiment. Now we discuss experimental signatures that characterize the $Z3OM$ state, and the origins of the superconductivity as a pair condensation out of it. 

\subsection{Anomalous vortex glass reinstated} 
In a previous paper\cite{Shi2025_anyon_delocalization}, we argued that if superconductivity is obtained through doping charge $2/3$ anyons, then in a disordered sample, the first superconductor that obtains is an Anomalous Vortex Glass (AVG) which has spontaneously nucleated vortices such that the net vorticity is zero. Subsequently we argued\cite{shi2026superconductivity} that for another anyonic route to superconductivity - obtained through pairing of fermionic partons - that the AVG will not result. We now examine this question for the route to superconductivity explored in this paper. 

Crucial to the answer is the fate of a localized anyon as the system transitions from the parent FQAH into the superconductor. We have already seen that the $1/3$ anyon will transmute into a $-h/2e \mod 3h/2e$ vortex when the $(2,2)$ boson condenses.  Now the arguments of Ref. \cite{Shi2025_anyon_delocalization} imply that in a disordered sample, the SC will be an AVG. Concomitantly~\cite{fisher1989vortex,fisher1991thermal,fisher1991vortex} there will be no finite-T BKT transition; there will be strongly non-linear IV characteristics etc which seem consistent with the observed superconductivity in Ref. \cite{Xu2025_SCdopeTMD}. 

\subsection{Fractional Josephson effect in SC-Z3OM -SC geometries} 
In moire materials, it may be possible to define gate-defined interfaces between different phases of the same device. Thus we can contemplate an SNS Josephson junction where the N region is the $Z3OM$. The presence of fractionally charge fermions in the N region will then have a distinctive signature on the Josephson effect (see Ref. \cite{senthil2001detecting} for a closely related discussion for detecting fractional charge with an emergent $Z_2$ gauge field) . It will allow charge-$2/3$ pairs to coherently tunnel between the two superconductors. Thus if voltage $V$ is applied across the junction, there will be an ac Josephson effect at frequency $\frac{2eV}{3h}$. Note however that if the N metal is replaced by a regular electronic Fermi liquid, then we will only get the usual charge-$2$ Josephson effect. This because the Fermi liquid metal only allows charge $2$ Cooper pairs to coherently tunnel through it. Thus the fractional Josephson effect is a probe of the physics of the normal metal and its relationship with the SC, rather than of the SC itself. 

\subsection{$2e/3$ shot noise in SC-Z3OM interfaces} 
As is well-known in a conventional superconductor-normal junction with a diffusive contact, the shot noise $S$ is doubled at bias voltages below the superconducting gap\cite{jehl2000detection,kozhevnikov2000observation}. Specifically $S = 4e F I$ where $I$ is the current through the junction and $F = 1/3$ is the Fano factor for a diffusive metal. For bias voltages above the gap, the noise becomes the usual $2eFI$. Consider now a gate defined interface between the Z3OM and the superconductor obtained by pair condensation of the $d$ fermions. Below the superconducting gap, we expect the shot noise $S = \frac{4e}{3} F I$  with $F = 1/3$ as before), which crosses over to $\frac{2e}{3} FI$ above the gap.  The extra factor of $1/3$ reflects the charge of the condensing boson. The physical picture is that in the $Z3OM$ region, the current is carried by the charge-$1/3$ fermions which diffuse, and get Andreev reflected at the interface from the $2/3$ condensate. 

Once again if the `normal' region is a normal electronic metal, even with the same superconductor, we will only get the usual answers for the noise without the extra factor of $1/3$. 

\subsection{Quantum oscillations} 
As metallic states with Fermi surfaces, both $Z3OM$ states will display quantum oscillations. 

Let us consider the $Z3OMs$ state first. As already noted, this has a single Fermi surface with area $\pi K_F^2 = (2\pi)^2 3 \delta \rho$ where $\delta\rho$ is the {\it deviation} of the electronic charge density from that in the parent $2/3$ FQAH state. In a small perpendicular magnetic field the $d$ fermions will form Landau levels which will lead to quantum oscillations. But due to the charge $1/3$,  the effective filling fraction of these fermions is $\nu^d_{eff} = 9 h\delta \rho/eB$. One factor of $3$ comes from writing the $d$-fermion density in terms of the electron density, and the other factor from the modified flux quantum $3h/e$. Thus we expect that in the $\delta \rho -B$ plane, there will resistivity minima forming a Landau fan  when $\nu^d_{eff}$ is an integer. The slope of these fans will be 9 times smaller than for regular electrons doped into the $2/3$ state. 

For the $Z3OMt$ state, there will again be Landau levels but now with a 3 fold valley degeneracy. 
In a field $B$, each species is at effective filling factor $\nu^\psi_{eff} = 3 h \delta \rho/eB$. The valley degeneracy means that the Landau fans occur when $\nu^\psi_{eff}$ is 3 times an integer. We will therefore have the usual Landau fan slope without a factor of $9$. If the valley degeneracy is split by disorder or other explicit translation breaking, then the Landau fan structure will be modified. 

Thus the Landau fan structure, if observed, can distinguish between the two $Z3OM$ states.

\subsection{ Single particle probes} 
A defining feature of Orthogonal Metals is that the low energy fermions that form the Fermi surface have no overlap with the electron. This will be reflected in a strong suppression of the single particle spectral weight at low frequencies. Consider for instance the tunneling conductance. This is determined by the Fourier transform of the local time dependent electron Greens function. The local $d$ fermion Green's function in $Z3OMs$ will decay $\sim 1/t$ at late times in $2d$. As the electron operator creates three $d$ fermions, the local electron Green's function will decay as $\sim 1/t^3$. For the tunneling conductance this translates into $G(V) \sim V^2$ at low bias voltage $V$. (The same result will also hold for $Z3OMt$). 

\subsection{Other measurements of fractionally charged fermions } 
 If thermal conductivity $\kappa$ in the metallic region can be measured accurately at low-$T$, we expect a Wiedemann-Franz law $\kappa_{xx}/T\sigma_{xx} = 9L_0$ where $L_0$ is the free electron Lorenz number. Thus the thermal conductivity will be much higher than might be naively expected from the electrical conductivity\footnote{Due to the high Hall conductivity, it is important that the ratio of conductivities rather than resistivities is considered.}. 

 \subsection{Charge order near vortices} 
 As noted above in the $Z3OMs$ state, the vortex is an $SU(3)$ fundamental (or $Z_3$ triplet if only that symmetry is present). Near a localized vortex we expect that that this triplet degree of freedom will freeze into some direction. Microscopically this corresponds to broken translation symmetry leading to period-$3$ charge order nucleated near the vortex core. In naive BCS theory applied to the fractional charge fermions, the vortex core might be expected to host the `normal' Orthogonal Metal state. Thus the possibility of charge order near the vortex core is an interesting effect that follows from the embedding of microscopic symmetries into the low energy Orthogonal Metal/BCS theory.

\section{Discussion}
The focus of this paper is on metallic ground states of the doped charge 1/3 anyon fluid that could be realized by doping the $2/3$ (or $1/3$) FQAH/FCI state. We discussed two distinct ``Orthogonal Metals"\cite{nandkishore2012orthogonal} with Fermi surfaces of sharply defined fermionic quasiparticles that carry fractional electric charge $1/3$ and are coupled to a discrete $Z_3$ gauge field. The discreteness of the gauge field implies that the low energy fractionally charged fermions near the fermi surface can be described by fermi liquid theory. Nevertheless there is no well defined electron quasiparticle at low energy and these Orthogonal Metals are not electronic Fermi liquids. A superconductor that arises through pairing of the fractionalized fermions is smoothly connected to a BCS superconductor, and depending on the pairing channel, can support any chiral central charge. 

These Orthogonal Metals are likely stabilized in the low doping regime when the effective Hamiltonian for the charge $1/3$ anyons is expected to have approximate $U(3)$ symmetry\cite{shi2026superconductivity}. This includes the `ideal' anyon gas limit of non-interacting anyons which may be the appropriate model at ultra-low doping when the inter-anyon spacing is much larger than the size of an anyon. 

We showed how the transport properties of the $Z3OM$ states might accommodate the high values of the resistivity seen in the normal metal proximate to the observed superconductivity in $tMoTe_2$. This leads us to suggest that a $Z3OM$ state might in fact already have been observed in $tMoTe_2$, and that it might be the normal state out of which superconductivity develops. We proposed a number of experimental tests of this idea. 

The two $Z3OM$ states will compete with other states that have been discussed in the literature (typically for generic interacting anyon fluids). These include a variety of superconducting states and a few metallic states (see the Table in Ref. ~\cite{shi2026superconductivity}). Of particular interest is a period-3 CDW metal that coexists with a background integer quantum Hall response. This state - with disorder localizing the quasiparticles at low density - offers a possible explanation\cite{Shi2025_anyon_delocalization,Nosov2025_plateau} of the Reentrant Integer Quantum Hall efect seen at large displacement field or large magnetic fields in $tMoTe_2$. Nevertheless, empirically there is a small pocket of metallic behavior at low displacement field and low magnetic fields that seems distinct. Our proposal is that this small pocket is one of the $Z3OM$ states.

\section*{Acknowledgment}

I thank Zhaoyu Han, Eslam Khalaf, Patrick Ledwith, Max Metlitski, Ashvin Vishwanath, and Xiaodong Xu for discussions. I am particularly thankful to Tingxin Li for many discussions on his experiments, and to Zhengyan Darius Shi for countless discussions and previous collaborations on this topic. This work was supported by the U.S. Department of Energy under Grant DE-SC0008739.

\bibliography{anyonOM}

@ARTICLE{nandkishore2012orthogonal,
       author = {{Nandkishore}, Rahul and {Metlitski}, Max A. and {Senthil}, T.},
        title = "{Orthogonal metals: The simplest non-Fermi liquids}",
      journal = {\prb},
         year = 2012,
        month = jul,
       volume = {86},
       number = {4},
          eid = {045128},
        pages = {045128},
          doi = {10.1103/PhysRevB.86.045128},
archivePrefix = {arXiv},
       eprint = {1201.5998},
 primaryClass = {cond-mat.str-el},
       adsurl = {https://ui.adsabs.harvard.edu/abs/2012PhRvB..86d5128N}
}

@article{liu2015characterization,
  title={Characterization of quasiholes in fractional Chern insulators},
  author={Liu, Zhao and Bhatt, Ravindra N and Regnault, Nicolas},
  journal={Physical Review B},
  volume={91},
  number={4},
  pages={045126},
  year={2015},
  publisher={APS}
}

@article{jehl2000detection,
  title={Detection of doubled shot noise in short normal-metal/superconductor junctions},
  author={Jehl, X and Sanquer, M and Calemczuk, R and Mailly, D},
  journal={Nature},
  volume={405},
  number={6782},
  pages={50--53},
  year={2000},
  publisher={Nature Publishing Group UK London}
}

@article{kozhevnikov2000observation,
  title={Observation of photon-assisted noise in a diffusive normal metal--superconductor junction},
  author={Kozhevnikov, AA and Schoelkopf, RJ and Prober, DE},
  journal={Physical Review Letters},
  volume={84},
  number={15},
  pages={3398},
  year={2000},
  publisher={APS}
}

@article{lu2012theory,
  title={Theory and classification of interacting integer topological phases in two dimensions: A Chern-Simons approach},
  author={Lu, Yuan-Ming and Vishwanath, Ashvin},
  journal={Physical Review B—Condensed Matter and Materials Physics},
  volume={86},
  number={12},
  pages={125119},
  year={2012},
  publisher={APS}
}

@article{senthil2013integer,
  title={Integer quantum hall effect for bosons},
  author={Senthil, T and Levin, Michael},
  journal={Physical review letters},
  volume={110},
  number={4},
  pages={046801},
  year={2013},
  publisher={APS}
}

@article{hsin2016level,
  title={Level/rank duality and Chern-Simons-matter theories},
  author={Hsin, Po-Shen and Seiberg, Nathan},
  journal={Journal of High Energy Physics},
  volume={2016},
  number={9},
  pages={1--30},
  year={2016},
  publisher={Springer}
}

@ARTICLE{Park2023_FQAH_TMD,
       author = {{Park}, Heonjoon and {Cai}, Jiaqi and {Anderson}, Eric and {Zhang}, Yinong and {Zhu}, Jiayi and {Liu}, Xiaoyu and {Wang}, Chong and {Holtzmann}, William and {Hu}, Chaowei and {Liu}, Zhaoyu and {Taniguchi}, Takashi and {Watanabe}, Kenji and {Chu}, Jiun-Haw and {Cao}, Ting and {Fu}, Liang and {Yao}, Wang and {Chang}, Cui-Zu and {Cobden}, David and {Xiao}, Di and {Xu}, Xiaodong},
        title = "{Observation of fractionally quantized anomalous Hall effect}",
      journal = {\nat},
         year = 2023,
        month = oct,
       volume = {622},
       number = {7981},
        pages = {74-79},
          doi = {10.1038/s41586-023-06536-0},
archivePrefix = {arXiv},
       eprint = {2308.02657},
 primaryClass = {cond-mat.mes-hall},
       adsurl = {https://ui.adsabs.harvard.edu/abs/2023Natur.622...74P}
}

@ARTICLE{Cai2023_FQAHTMD,
       author = {{Cai}, Jiaqi and {Anderson}, Eric and {Wang}, Chong and {Zhang}, Xiaowei and {Liu}, Xiaoyu and {Holtzmann}, William and {Zhang}, Yinong and {Fan}, Fengren and {Taniguchi}, Takashi and {Watanabe}, Kenji and {Ran}, Ying and {Cao}, Ting and {Fu}, Liang and {Xiao}, Di and {Yao}, Wang and {Xu}, Xiaodong},
        title = "{Signatures of fractional quantum anomalous Hall states in twisted MoTe$_{2}$}",
      journal = {\nat},
         year = 2023,
        month = oct,
       volume = {622},
       number = {7981},
        pages = {63-68},
          doi = {10.1038/s41586-023-06289-w},
archivePrefix = {arXiv},
       eprint = {2304.08470},
 primaryClass = {cond-mat.mes-hall},
       adsurl = {https://ui.adsabs.harvard.edu/abs/2023Natur.622...63C}
}

@ARTICLE{Xu2023_FQAHTMD,
       author = {{Xu}, Fan and {Sun}, Zheng and {Jia}, Tongtong and {Liu}, Chang and {Xu}, Cheng and {Li}, Chushan and {Gu}, Yu and {Watanabe}, Kenji and {Taniguchi}, Takashi and {Tong}, Bingbing and {Jia}, Jinfeng and {Shi}, Zhiwen and {Jiang}, Shengwei and {Zhang}, Yang and {Liu}, Xiaoxue and {Li}, Tingxin},
        title = "{Observation of Integer and Fractional Quantum Anomalous Hall Effects in Twisted Bilayer MoTe$_{2}$}",
      journal = {Physical Review X},
         year = 2023,
        month = jul,
       volume = {13},
       number = {3},
          eid = {031037},
        pages = {031037},
          doi = {10.1103/PhysRevX.13.031037},
archivePrefix = {arXiv},
       eprint = {2308.06177},
 primaryClass = {cond-mat.mes-hall},
       adsurl = {https://ui.adsabs.harvard.edu/abs/2023PhRvX..13c1037X}
}

@ARTICLE{Lu2023_FQAHPenta,
       author = {{Lu}, Zhengguang and {Han}, Tonghang and {Yao}, Yuxuan and {Reddy}, Aidan P. and {Yang}, Jixiang and {Seo}, Junseok and {Watanabe}, Kenji and {Taniguchi}, Takashi and {Fu}, Liang and {Ju}, Long},
        title = "{Fractional quantum anomalous Hall effect in multilayer graphene}",
      journal = {\nat},
         year = 2024,
        month = feb,
       volume = {626},
       number = {8000},
        pages = {759-764},
          doi = {10.1038/s41586-023-07010-7},
archivePrefix = {arXiv},
       eprint = {2309.17436},
 primaryClass = {cond-mat.mes-hall},
       adsurl = {https://ui.adsabs.harvard.edu/abs/2024Natur.626..759L}
}

@ARTICLE{Lu2025_EQAH,
       author = {{Lu}, Zhengguang and {Han}, Tonghang and {Yao}, Yuxuan and {Hadjri}, Zach and {Yang}, Jixiang and {Seo}, Junseok and {Shi}, Lihan and {Ye}, Shenyong and {Watanabe}, Kenji and {Taniguchi}, Takashi and {Ju}, Long},
        title = "{Extended quantum anomalous Hall states in graphene/hBN moir{\'e} superlattices}",
      journal = {\nat},
         year = 2025,
        month = jan,
       volume = {637},
       number = {8048},
        pages = {1090-1095},
          doi = {10.1038/s41586-024-08470-1},
archivePrefix = {arXiv},
       eprint = {2408.10203},
 primaryClass = {cond-mat.mes-hall},
       adsurl = {https://ui.adsabs.harvard.edu/abs/2025Natur.637.1090L}
}

@article{Zeng2023_FQAHTMD,
	author = {Zeng, Yihang and Xia, Zhengchao and Kang, Kaifei and Zhu, Jiacheng and Kn{\"u}ppel, Patrick and Vaswani, Chirag and Watanabe, Kenji and Taniguchi, Takashi and Mak, Kin Fai and Shan, Jie},
	da = {2023/10/01},
	doi = {10.1038/s41586-023-06452-3},
	id = {Zeng2023},
	isbn = {1476-4687},
	journal = {Nature},
	number = {7981},
	pages = {69--73},
	title = {Thermodynamic evidence of fractional Chern insulator in moir{\'e}MoTe2},
	ty = {JOUR},
	url = {https://doi.org/10.1038/s41586-023-06452-3},
	volume = {622},
	year = {2023}}

@ARTICLE{Xu2025_SCdopeTMD,
       author = {{Xu}, Fan and {Sun}, Zheng and {Li}, Jiayi and {Zheng}, Ce and {Xu}, Cheng and {Gao}, Jingjing and {Jia}, Tongtong and {Watanabe}, Kenji and {Taniguchi}, Takashi and {Tong}, Bingbing and {Lu}, Li and {Jia}, Jinfeng and {Shi}, Zhiwen and {Jiang}, Shengwei and {Zhang}, Yuanbo and {Zhang}, Yang and {Lei}, Shiming and {Liu}, Xiaoxue and {Li}, Tingxin},
        title = "{Signatures of unconventional superconductivity near reentrant and fractional quantum anomalous Hall insulators}",
      journal = {arXiv e-prints},
         year = 2025,
        month = apr,
          eid = {arXiv:2504.06972},
        pages = {arXiv:2504.06972},
          doi = {10.48550/arXiv.2504.06972},
archivePrefix = {arXiv},
       eprint = {2504.06972},
 primaryClass = {cond-mat.mes-hall},
       adsurl = {https://ui.adsabs.harvard.edu/abs/2025arXiv250406972X}
}

@article{Laughlin1988_anyonSC,
  title = {Superconducting Ground State of Noninteracting Particles Obeying Fractional Statistics},
  author = {Laughlin, R. B.},
  journal = {Phys. Rev. Lett.},
  volume = {60},
  issue = {25},
  pages = {2677--2680},
  numpages = {0},
  year = {1988},
  month = {Jun},
  publisher = {American Physical Society},
  doi = {10.1103/PhysRevLett.60.2677},
  url = {https://link.aps.org/doi/10.1103/PhysRevLett.60.2677}
}

@article{Fetter1989_anyonSC_RPA,
  title = {Random-phase approximation in the fractional-statistics gas},
  author = {Fetter, A. L. and Hanna, C. B. and Laughlin, R. B.},
  journal = {Phys. Rev. B},
  volume = {39},
  issue = {13},
  pages = {9679--9681},
  numpages = {0},
  year = {1989},
  month = {May},
  publisher = {American Physical Society},
  doi = {10.1103/PhysRevB.39.9679},
  url = {https://link.aps.org/doi/10.1103/PhysRevB.39.9679}
}

@article{Lee1989_anyonSC,
  title = {Anyon superconductivity and the fractional quantum Hall effect},
  author = {Lee, Dung-Hai and Fisher, Matthew P. A.},
  journal = {Phys. Rev. Lett.},
  volume = {63},
  issue = {8},
  pages = {903--906},
  numpages = {0},
  year = {1989},
  month = {Aug},
  publisher = {American Physical Society},
  doi = {10.1103/PhysRevLett.63.903},
  url = {https://link.aps.org/doi/10.1103/PhysRevLett.63.903}
}

@ARTICLE{Chen1989_anyonSC,
       author = {{Chen}, Yi-Hong and {Wilczek}, Frank and {Witten}, Edward and {Halperin}, Bertrand I.},
        title = "{On Anyon Superconductivity}",
      journal = {International Journal of Modern Physics A},
         year = 1989,
        month = jan,
       volume = {4},
       number = {15},
        pages = {3983},
          doi = {10.1142/S0217751X89001631},
       adsurl = {https://ui.adsabs.harvard.edu/abs/1989IJMPA...4.3983C}
}

@article{Wen1990_anyonSC,
  title = {Compressibility and superfluidity in the fractional-statistics liquid},
  author = {Wen, X. G. and Zee, A.},
  journal = {Phys. Rev. B},
  volume = {41},
  issue = {1},
  pages = {240--253},
  numpages = {0},
  year = {1990},
  month = {Jan},
  publisher = {American Physical Society},
  doi = {10.1103/PhysRevB.41.240},
  url = {https://link.aps.org/doi/10.1103/PhysRevB.41.240}
}

@article{fisher1989vortex,
  title = {Vortex-glass superconductivity: A possible new phase in bulk high-${\mathrm{T}}_{\mathrm{c}}$ oxides},
  author = {Fisher, Matthew P. A.},
  journal = {Phys. Rev. Lett.},
  volume = {62},
  issue = {12},
  pages = {1415--1418},
  numpages = {0},
  year = {1989},
  month = {Mar},
  publisher = {American Physical Society},
  doi = {10.1103/PhysRevLett.62.1415},
  url = {https://link.aps.org/doi/10.1103/PhysRevLett.62.1415}
}

@article{fisher1991thermal,
  title = {Thermal fluctuations, quenched disorder, phase transitions, and transport in type-II superconductors},
  author = {Fisher, Daniel S. and Fisher, Matthew P. A. and Huse, David A.},
  journal = {Phys. Rev. B},
  volume = {43},
  issue = {1},
  pages = {130--159},
  numpages = {0},
  year = {1991},
  month = {Jan},
  publisher = {American Physical Society},
  doi = {10.1103/PhysRevB.43.130},
  url = {https://link.aps.org/doi/10.1103/PhysRevB.43.130}
}

@article{fisher1991vortex,
  title = {Vortex variable-range-hopping resistivity in superconducting films},
  author = {Fisher, Matthew P. A. and Tokuyasu, T. A. and Young, A. P.},
  journal = {Phys. Rev. Lett.},
  volume = {66},
  issue = {22},
  pages = {2931--2934},
  numpages = {0},
  year = {1991},
  month = {Jun},
  publisher = {American Physical Society},
  doi = {10.1103/PhysRevLett.66.2931},
  url = {https://link.aps.org/doi/10.1103/PhysRevLett.66.2931}
}

@article{senthil2001detecting,
  title={Detecting fractions of electrons in the high-T c cuprates},
  author={Senthil, T and Fisher, Matthew PA},
  journal={Physical Review B},
  volume={64},
  number={21},
  pages={214511},
  year={2001},
  publisher={APS}
}

@ARTICLE{Tang2013_anyonSC,
       author = {{Tang}, Evelyn and {Wen}, Xiao-Gang},
        title = "{Superconductivity with intrinsic topological order induced by pure Coulomb interaction and time-reversal symmetry breaking}",
      journal = {\prb},
         year = 2013,
        month = nov,
       volume = {88},
       number = {19},
          eid = {195117},
        pages = {195117},
          doi = {10.1103/PhysRevB.88.195117},
archivePrefix = {arXiv},
       eprint = {1306.1528},
 primaryClass = {cond-mat.str-el},
       adsurl = {https://ui.adsabs.harvard.edu/abs/2013PhRvB..88s5117T}
}

@ARTICLE{Shi2024_doping,
       author = {{Shi}, Zhengyan Darius and {Senthil}, T.},
        title = "{Doping a Fractional Quantum Anomalous Hall Insulator}",
      journal = {Physical Review X},
         year = 2025,
        month = jul,
       volume = {15},
       number = {3},
          eid = {031069},
        pages = {031069},
          doi = {10.1103/kcm5-hx56},
archivePrefix = {arXiv},
       eprint = {2409.20567},
 primaryClass = {cond-mat.str-el},
       adsurl = {https://ui.adsabs.harvard.edu/abs/2025PhRvX..15c1069S}
}

@ARTICLE{Kim2024_anyonSC,
       author = {{Kim}, Minho and {Timmel}, Abigail and {Ju}, Long and {Wen}, Xiao-Gang},
        title = "{Topological chiral superconductivity beyond pairing in a Fermi liquid}",
      journal = {\prb},
         year = 2025,
        month = jan,
       volume = {111},
       number = {1},
          eid = {014508},
        pages = {014508},
          doi = {10.1103/PhysRevB.111.014508},
archivePrefix = {arXiv},
       eprint = {2409.18067},
 primaryClass = {cond-mat.str-el},
       adsurl = {https://ui.adsabs.harvard.edu/abs/2025PhRvB.111a4508K}
}

@ARTICLE{Divic2024_HofHubb,
       author = {{Divic}, Stefan and {Cr{\'e}pel}, Valentin and {Soejima}, Tomohiro and {Song}, Xue-Yang and {Millis}, Andrew J. and {Zaletel}, Michael P. and {Vishwanath}, Ashvin},
        title = "{Anyon superconductivity from topological criticality in a Hofstadter-Hubbard model}",
      journal = {Proceedings of the National Academy of Science},
         year = 2025,
        month = aug,
       volume = {122},
       number = {33},
          eid = {e2426680122},
        pages = {e2426680122},
          doi = {10.1073/pnas.2426680122},
archivePrefix = {arXiv},
       eprint = {2410.18175},
 primaryClass = {cond-mat.str-el},
       adsurl = {https://ui.adsabs.harvard.edu/abs/2025PNAS..12226680D}
}

@ARTICLE{Shi2025_dopeMR,
       author = {{Darius Shi}, Zhengyan and {Zhang}, Carolyn and {Senthil}, T.},
        title = "{Doping lattice non-abelian quantum Hall states}",
      journal = {arXiv e-prints},
         year = 2025,
        month = may,
          eid = {arXiv:2505.02893},
        pages = {arXiv:2505.02893},
          doi = {10.48550/arXiv.2505.02893},
archivePrefix = {arXiv},
       eprint = {2505.02893},
 primaryClass = {cond-mat.str-el},
       adsurl = {https://ui.adsabs.harvard.edu/abs/2025arXiv250502893D}
}

@ARTICLE{Shi2025_anyon_delocalization,
       author = {{Darius Shi}, Zhengyan and {Senthil}, T.},
        title = "{Anyon delocalization transitions out of a disordered FQAH insulator}",
      journal = {arXiv e-prints},
         year = 2025,
        month = jun,
          eid = {arXiv:2506.02128},
        pages = {arXiv:2506.02128},
          doi = {10.48550/arXiv.2506.02128},
archivePrefix = {arXiv},
       eprint = {2506.02128},
 primaryClass = {cond-mat.str-el},
       adsurl = {https://ui.adsabs.harvard.edu/abs/2025arXiv250602128D}
}

@ARTICLE{Nosov2025_plateau,
       author = {{Nosov}, Pavel A. and {Han}, Zhaoyu and {Khalaf}, Eslam},
        title = "{Anyon superconductivity and plateau transitions in doped fractional quantum anomalous Hall insulators}",
      journal = {arXiv e-prints},
         year = 2025,
        month = jun,
          eid = {arXiv:2506.02108},
        pages = {arXiv:2506.02108},
          doi = {10.48550/arXiv.2506.02108},
archivePrefix = {arXiv},
       eprint = {2506.02108},
 primaryClass = {cond-mat.str-el},
       adsurl = {https://ui.adsabs.harvard.edu/abs/2025arXiv250602108N}
}

@ARTICLE{Pichler2025_anyonSC,
       author = {{Pichler}, Fabian and {Kuhlenkamp}, Clemens and {Knap}, Michael and {Vishwanath}, Ashvin},
        title = "{Microscopic Mechanism of Anyon Superconductivity Emerging from Fractional Chern Insulators}",
      journal = {arXiv e-prints},
         year = 2025,
        month = jun,
          eid = {arXiv:2506.08000},
        pages = {arXiv:2506.08000},
          doi = {10.48550/arXiv.2506.08000},
archivePrefix = {arXiv},
       eprint = {2506.08000},
 primaryClass = {cond-mat.str-el},
       adsurl = {https://ui.adsabs.harvard.edu/abs/2025arXiv250608000P}
}

@ARTICLE{Han2025_anyonexciton,
       author = {{Han}, Zhaoyu and {Wang}, Taige and {Dong}, Zhihuan and {Zaletel}, Michael P. and {Vishwanath}, Ashvin},
        title = "{Anyon superfluidity of excitons in quantum Hall bilayers}",
      journal = {arXiv e-prints},
         year = 2025,
        month = aug,
          eid = {arXiv:2508.14894},
        pages = {arXiv:2508.14894},
          doi = {10.48550/arXiv.2508.14894},
archivePrefix = {arXiv},
       eprint = {2508.14894},
 primaryClass = {cond-mat.str-el},
       adsurl = {https://ui.adsabs.harvard.edu/abs/2025arXiv250814894H}
}

@ARTICLE{Nakajima2025_thermo_anyon,
       author = {{Nakajima}, Yuto and {Mehta}, Umang and {Goldman}, Hart},
        title = "{Thermodynamics of dilute anyon gases from fusion constraints}",
      journal = {arXiv e-prints},
         year = 2025,
        month = aug,
          eid = {arXiv:2508.14961},
        pages = {arXiv:2508.14961},
          doi = {10.48550/arXiv.2508.14961},
archivePrefix = {arXiv},
       eprint = {2508.14961},
 primaryClass = {cond-mat.str-el},
       adsurl = {https://ui.adsabs.harvard.edu/abs/2025arXiv250814961N}
}

@ARTICLE{Kuhlenkamp2025_HofHubb,
       author = {{Kuhlenkamp}, Clemens and {Divic}, Stefan and {Zaletel}, Michael P. and {Soejima}, Tomohiro and {Vishwanath}, Ashvin},
        title = "{Robust superconductivity upon doping chiral spin liquid and Chern insulators in a Hubbard-Hofstadter model}",
      journal = {arXiv e-prints},
         year = 2025,
        month = sep,
          eid = {arXiv:2509.02675},
        pages = {arXiv:2509.02675},
          doi = {10.48550/arXiv.2509.02675},
archivePrefix = {arXiv},
       eprint = {2509.02675},
 primaryClass = {cond-mat.str-el},
       adsurl = {https://ui.adsabs.harvard.edu/abs/2025arXiv250902675K}
}

@ARTICLE{lotrivc2026phases,
       author = {{Lotri{$\v{c}$}}, Tev{$\v{z}$} and {Simon}, Steven H.},
        title = "{Phases of itinerant anyons in Laughlin's quantum Hall states on a lattice}",
      journal = {arXiv e-prints},
         year = 2026,
        month = mar,
          eid = {arXiv:2603.22389},
        pages = {arXiv:2603.22389},
          doi = {10.48550/arXiv.2603.22389},
archivePrefix = {arXiv},
       eprint = {2603.22389},
 primaryClass = {cond-mat.str-el},
       adsurl = {https://ui.adsabs.harvard.edu/abs/2026arXiv260322389L}
}

@ARTICLE{Shi2025_nonAbelian_TSC,
       author = {{Darius Shi}, Zhengyan and {Senthil}, T.},
        title = "{Non-Abelian topological superconductivity from melting Abelian fractional Chern insulators}",
      journal = {arXiv e-prints},
         year = 2025,
        month = dec,
          eid = {arXiv:2512.17996},
        pages = {arXiv:2512.17996},
          doi = {10.48550/arXiv.2512.17996},
archivePrefix = {arXiv},
       eprint = {2512.17996},
 primaryClass = {cond-mat.str-el},
       adsurl = {https://ui.adsabs.harvard.edu/abs/2025arXiv251217996D}
}

@ARTICLE{Fan2026_weakpairing_SC,
       author = {{Fan}, Zheng-Duo and {Vishwanath}, Ashvin and {Wang}, Zijian},
        title = "{Hidden weak-pairing superconductivity of non-interacting anyons obeying $\frac{1}{3}$ statistics}",
      journal = {arXiv e-prints},
         year = 2026,
        month = may,
          eid = {arXiv:2605.19036},
        pages = {arXiv:2605.19036},
          doi = {10.48550/arXiv.2605.19036},
archivePrefix = {arXiv},
       eprint = {2605.19036},
 primaryClass = {cond-mat.str-el},
       adsurl = {https://ui.adsabs.harvard.edu/abs/2026arXiv260519036F}
}

@ARTICLE{Wang2026_U3anyonSC,
       author = {{Wang}, Taige},
        title = "{Topological superconductivity from Abelian fractional Chern insulators}",
      journal = {arXiv e-prints},
         year = 2026,
        month = may,
          eid = {arXiv:2605.29034},
        pages = {arXiv:2605.29034},
          doi = {10.48550/arXiv.2605.29034},
archivePrefix = {arXiv},
       eprint = {2605.29034},
 primaryClass = {cond-mat.str-el},
       adsurl = {https://ui.adsabs.harvard.edu/abs/2026arXiv260529034W}
}

@article{shi2026superconductivity,
  title={Superconductivity and non-Fermi liquid metals in a charge-1/3 anyon fluid},
  author={Shi, Zhengyan Darius and Senthil, T},
  journal={arXiv preprint arXiv:2606.20403},
  year={2026}
}

@ARTICLE{Yan2025_anyondisp,
       author = {{Yan}, Zihan and {Li}, Qingchen and {Soejima}, Tomohiro and {Khalaf}, Eslam},
        title = "{Anyon Dispersion in Aharonov-Casher Bands and Implications for Twisted MoTe${}_2$}",
      journal = {arXiv e-prints},
         year = 2025,
        month = dec,
          eid = {arXiv:2512.15863},
        pages = {arXiv:2512.15863},
          doi = {10.48550/arXiv.2512.15863},
archivePrefix = {arXiv},
       eprint = {2512.15863},
 primaryClass = {cond-mat.str-el},
       adsurl = {https://ui.adsabs.harvard.edu/abs/2025arXiv251215863Y}
}
\onecolumngrid
\newpage 
\appendix
\section{Z3OMs from Jain composite fermions}
\label{app: z3omsJain}
As explained in previous papers\cite{Shi2024_doping,shi2026superconductivity}, the Jain composite fermion theory for doped $1/3$ charges near the $1/3$ FQAH state takes the form 
\begin{equation} 
{\cal L} = \sum_{I = 1}^3 {\cal L} [f_I, a] + \frac{1}{4\pi} ada + 2 CS[g] - \frac{2}{4\pi} bdb + \frac{1}{2\pi} b d(A-a) 
\end{equation} 
Here $f_I$ are fields that represent Jain composite fermions. At a small doping density, the $f_I$ see a mean magnetic field corresponding to Landau level filling $-3/2$. In Ref. ~\cite{shi2026superconductivity} we showed how the $Z3OMt$ state can arise from this theory. Here we show how to also obtain the $Z3OMs$ state from this theory. To that end we use a parton description $f_I = \hat{\Phi}_I d$ in terms of a an $SU(3)$ fundamental boson $\hat{\Phi}_I$ and an $SU(3)$ singlet fermion $d$. We let $\hat{\Phi}_i$ see the magnetic field while the $d$ forms a Fermi surface. Then $\hat{\Phi}_I$ is at Landau level filling $-3/2$. A $U(3)$ symmetric  fractional quantum Hall state for bosons exists at this filling: it is simply a boson integer quantum Hall state with $\sigma_{xy} = -2$ stacked with a Laughlin $\sigma_{xy} = 1/2$ state. To see that this topological order can be enriched with $U(3)$ symmetry, note that the anyon content of the theory is just ${1,s}$ where $s$ is a semion that carries charge $1/2$ under the global $U(1)$ symmetry. We know that the fusion of two semions must include $\hat{\Phi}_I$  which transforms in the $\bs{3}$ representation of $SU(3)$. If the semion transforms in the $\bs{\bar{3}}$ rep, then as $\bs{\bar{3}} \times \bs{\bar{3}} = \bs{3} + \bar{\bs{6}}$, we can obtain $\hat{\Phi}_I$ through the fusion of two semions\footnote{A specific construction of this $U(3)$ enriched topological order can be obtained by writing $\hat{\Phi}_I = \Psi_I \chi$, letting $\Psi_I$ fill $3$ Landau levels and $\chi$ fill a $-1$ Landau level.}.

Placing $\hat{\Phi}_I$ in this quantum Hall state, and repeating the same steps as in Ref~\cite{shi2026superconductivity}, we get the $Z3OMs$ theory of Eqn. \ref{eq: z3oms}. 

Understanding the path to these states from Jain composite fermions is likely a useful guide to energetics given the expectation that these composite fermions provide a useful description of the short distance physics of anyons in most fractional quantum Hall states, including those observed at zero field.

\section{Superconductivity from pairing charge $1/3$ fermions in the $Z3OM$: formal description}
\label{app: SC} 
Here we supplement the physical arguments in the main text with a formal field theoretic description of the superconducting states obtained by condensing pair of the low energy charge $1/3$ fermions of either of the two $Z3OM$ states. The treatment below is identical to both $Z3OMs$ and $Z3OMt$. Hence, for concreteness, we focus on the $Z3OMs$ below, and worked with the doped $1/3$ state (Eqn. \ref{eq: z3oms}). 

We introduce a pair field $\psi \sim dd$ that carries charge-$2$ under the spin$_c$ connection $a$.  The Lagrangian is then (schematically) 
\begin{equation} 
{\cal L} = {\cal L}[\psi, 2a] + \bar{\psi} dd  + \cdots  + {\cal L}[d,a] + 3CS[a,g] + \frac{1}{2\pi} bd (A - 3a)
\end{equation} 
Assume that the pair condensation fully gaps the superconductor and that there are $n$ Majorana fermion edge modes that result. Including the contribution of the $CS[a,g]$ term, the total chiral central charge will then be $2c_- = n + 6$. The low energy theory of the paired state is then 
\begin{equation}
    {\cal L} = \frac{2}{2\pi} cda + \frac{3}{4\pi} ada + \frac{1}{2\pi} bd (A - 3a) + (n+6) CS[g] 
\end{equation}
where $c$ is an ordinary $U(1)$ gauge field that describes the Higgsing of $2a$ introduced by the pair condensation. To expose the physics, shift $b = \hat{b} +c$ with $\hat{b}$ another $U(1)$ gauge field to get 
\begin{equation} 
{\cal L} = \frac{1}{2\pi} cda +  \frac{3}{4\pi} ada - \frac{3}{2\pi }\hat{b} da + \frac{1}{2\pi} \hat{b} d (A - 3a) - \frac{1}{2\pi} Adb + (n+6) CS[g] 
\end{equation} 
Now the $c$ field becomes a Lagrange multiplier that sets $a = A$. Thus we get 
\begin{equation}
{\cal L} = -\frac{2}{2\pi} Ad\hat{b} + \frac{3}{4\pi} AdA + (n+6)CS_g 
\end{equation} 
This describes an ordinary charge-$2$ electronic superconductor with $(n+6)$ Majorana edge modes. 

In application to the doped $2/3$ state we need to particle-hole conjugate the above and add the contribution of the filled Chern band. This gives a charge-$2$ superconductor with chiral central charge $c_- = -2 - n/2 $. 

In the strong pairing limit we will have $n = 0$ and $c_- = -2$ (the same as that obtained most naturally by directly doping charge $2/3$ anyons) while weak pairing in a state with angular momentum $l$ for the $\psi$ will lead to $n = l$ (with $l$ odd). Thus we can get any half-integer chiral central charge.

\section{N-species anyon fluids with $\pi/N$ statistics}
We now briefly consider a slight generalization to fluids of $N$ flavors of anyons with $\pi/N$ statistics. We start with the generalization of Eq. \ref{eq: bosonrep}: 
\begin{equation} 
\label{eq: bosonrepN}
{\cal S} = \int d^2x d\tau \sum_{I =1}^N \bar{\Phi}_I \left(i\partial_t + b_0 + \frac{1}{2m}(-i \vec \nabla - \vec b)^2 \right)\Phi_I - \frac{N}{4\pi} bdb + \frac{1}{2\pi} Adb + {\cal S}_{int}
\end{equation} 
We will restrict to the case with full global $U(N)$ symmetry. 

For any $N$, a possible ground state is that the $\Phi_I $ condenses into a superfluid that spontaneously breaks the $SU(N)$ symmetry.  The presence of the mean magnetic field implies that a vortex lattice will form. The quantization of the $b$ flux at each vortex implies that this state is an electron Wigner crystal. With increasing $N$, this may well be energetically favored. 

As alternate possibilities, the steps in the main text leading to the Orthogonal Metal states can clearly be repeated for any $N$. Briefly, we write $\Phi_I = \psi_I d$ and either let $d$ form an integer quantum Hall state with $N$ filled Landau levels and $\psi_I$ form Fermi surfaces, or let $d$ form a Fermi surface and let $\psi_I$ each fill a Landau level. We then get two distinct $ZNOM$ metallic states where Fermi surface(s) of sharply defined charge $1/N$ fermionic quasiparticles are coupled to a discrete $Z_N$ gauge field. Either of these states preserves the global $U(N)$ symmetry. 

For $N$ even, there is another natural competing state. As the bosons $\Phi_I$ are at an even integer filling $N$, they can form a $U(N)$ symmetric boson integer quantum Hall state\cite{lu2012theory,senthil2013integer,hsin2016level}. The Chern-Simons term for $b$ is then canceled and we get a charge 1 superfluid. Note that for even $N$ the theory is bosonic ({\it i.e} all local operators are bosons), and a charge 1 superfluid is allowed. 

This option is not possible for odd $N$ as bosons at odd integer filling cannot form a boson integer quantum Hall state. 

For the future it may be worthwhile to explore the model in the large-$N$ limit to see whether the true ground state can be determined in a controlled analytic calculation. 

\end{document}